\documentclass[
twocolumn,
longbibliography,
superscriptaddress,
amsmath,amssymb,
aps,
pra
]{revtex4-2}

\usepackage{graphicx}
\usepackage{dcolumn}
\usepackage{bm}
\usepackage{ulem}
\usepackage{hyperref}
\usepackage{color}
\newcommand{\wzy}[1]{\textcolor{black}{#1}}
\newcommand{\cc}[1]{\textcolor{black}{#1}}

\newcommand{\ljy}[1]{\textcolor{black}{#1}}

\begin{document}

\title{Fast adiabatic preparation of multi-squeezed states by jumping along the path}

\author{Chuan Chen}
\affiliation {Key Laboratory of Atomic and Subatomic Structure and Quantum Control (Ministry of Education), Guangdong Basic Research Center of Excellence for Structure and Fundamental Interactions of Matter, School of Physics, South China Normal University, Guangzhou 510006, China}
\author{Jian-Yu Lu}
\affiliation {Key Laboratory of Atomic and Subatomic Structure and Quantum Control (Ministry of Education), Guangdong Basic Research Center of Excellence for Structure and Fundamental Interactions of Matter, School of Physics, South China Normal University, Guangzhou 510006, China}
\author{Xu-Yang Chen}
\email{2022010138@m.scnu.edu.cn}
\affiliation {Key Laboratory of Atomic and Subatomic Structure and Quantum Control (Ministry of Education), Guangdong Basic Research Center of Excellence for Structure and Fundamental Interactions of Matter, School of Physics, South China Normal University, Guangzhou 510006, China}
\author{Zhen-Yu Wang}
\email{zhenyu.wang@m.scnu.edu.cn}
\affiliation {Key Laboratory of Atomic and Subatomic Structure and Quantum Control (Ministry of Education), Guangdong Basic Research Center of Excellence for Structure and Fundamental Interactions of Matter, School of Physics, South China Normal University, Guangzhou 510006, China} 
\affiliation {Guangdong Provincial Key Laboratory of Quantum Engineering and Quantum Materials, Guangdong-Hong Kong Joint Laboratory of Quantum Matter, South China Normal University, Guangzhou 510006, China}

\begin{abstract}

Multi-squeezed states, also known as generalized squeezed states, are valuable quantum non-Gaussian resources, because they \wzy{can} feature non-classical properties such as large phase-space Wigner negativities. In this work, we introduce a novel shortcuts to adiabaticity (STA) method for the fast preparation of multi-squeezed states. In contrast to previous STA methods, which rely on the use of counterdiabatic control to suppress unwanted non-adiabatic effects, our method simplifies the process and accelerates state preparation by selecting an appropriate sampling along a quantum evolution path. We demonstrate the high-fidelity and fast preparation of multi-squeezed states, as well as hybrid entangled states between a bosonic mode and a qubit.

\end{abstract}

\maketitle

\section{Introduction}

The quantum adiabatic control is a fundamental concept in quantum mechanics~\cite{messiah1961book,berry1984quantal,albash2018adiabatic}, which implies that a physical system can be kept in its instantaneous eigenstates by applying a slowly varying control field. This facilitates various robust preparations of quantum states, such as coherent and squeezed states~\cite{leroux2018enhancing,zheng2012squeezing,chen2019squeezing,andersen2016squeezed}.

However, the slow evolution, which is required in the traditional adiabatic approximation~\cite{messiah1961book}, makes the quantum system more susceptible to decoherence. To reduce the effects of decoherence, shortcuts to adiabaticity (STA) methods~\cite{demirplak2003adiabatic,berry2009transitionless,chen2010shortcut,
torrontegui2013sta,odelin2019shortcuts,bason2012high,baksic2016sta,zhou2017accelerated,du2016experimental,song2021robust} have been developed to accelerate the adiabatic evolution. By the use of counterdiabatic driving fields to cancel out the non-adiabatic effects of the original Hamiltonian, STA schemes achieve the same outcome as traditional quantum adiabatic methods in a much shorter time.

Despite their great success, these STA methods~\cite{demirplak2003adiabatic,berry2009transitionless,chen2010shortcut,
torrontegui2013sta,odelin2019shortcuts,bason2012high,baksic2016sta,du2016experimental,zhou2017accelerated,song2021robust} 
necessitate counterdiabatic driving, which is not always feasible or practical, due to the potential requirement of experimental resources that are hard to access~\cite{bason2012high,baksic2016sta,zhou2017accelerated}. Furthermore, the counterdiabatic control fields also change the eigenstates of the original Hamiltonian and potentially introduce additional control errors, which might cause the quantum system to deviate from the intended evolution path. These factors make it difficult to design a robust STA method that can withstand control errors. 

To address the problems mentioned above, shortcut to adiabaticity by modulation (STAM) method has been proposed~\cite{liu2022shortcuts}. This STAM is based on the necessary and sufficient condition of quantum adiabatic \wzy{evolution~\cite{wang2016necessary,xu2019geodesics}}. It eliminates the non-adiabatic effects by dynamically adjusting the Hamiltonian parameters, without the need of auxiliary driving. Therefore, STAM method provides a new approach to construct parameterized Hamiltonians for adiabatic control~\cite{zheng2022accelerated,gong2023jumping} and new insights to manipulate quantum systems \wzy{and sensing~\cite{xu2024control,zeng2024wide}}.

Here we develop a STAM method to prepare multi-squeezed states~\cite{fisher1984impossibility,braunstein1987squeezing,
hillery1984nonlinear,vourdas1992squeezing,hillery1990divergence,vraunstein1990statistics}. \cc{Multi-squeezed states, the higher-order generalization of coherent (first order) and squeezed states (second order)~\cite{glauber1963coherent,zhang1990coherentReview,walls1963squeezed}, exhibit remarkable non-Gaussian properties for third order and beyond, such as large phase-space Wigner negativities, interference, and enhanced squeezing~\cite{fisher1984impossibility,braunstein1987squeezing,
hillery1984nonlinear,vourdas1992squeezing,hillery1990divergence,vraunstein1990statistics}. These unique characteristics have made them considered as strictly quantifiable and valuable non-Gaussian resources in quantum information technology~\cite{mcconnell2022multisqueezed,banaszek1997interference,chang2020experiment,albarelli2018wigner,zheng2021conversion,wenlong2021circuit}, despite initial beliefs that they were physically impossible~\cite{fisher1984impossibility}. However, subsequent theoretical analyses demonstrated the generation of multi-squeezed states via spontaneous parametric down-conversion (SPDC)~\cite{braunstein1987squeezing,hillery1990divergence}, with recent experimental realization for the third order of multi-squeezed states~\cite{chang2020experiment}. }

The paper is organized as follows. \wzy{In Sec.~\ref{sec:2}, we construct our STAM method and show superior performance of our method over traditional adiabatic control in the preparation of multi-squeezed states. In Sec.~\ref{sec:3}, we extend our method to the preparation of hybrid entangled states. We draw our conclusions in Sec.~\ref{sec:conclusion}.}

\begin{figure}
    \centering
    \includegraphics[width=8cm]{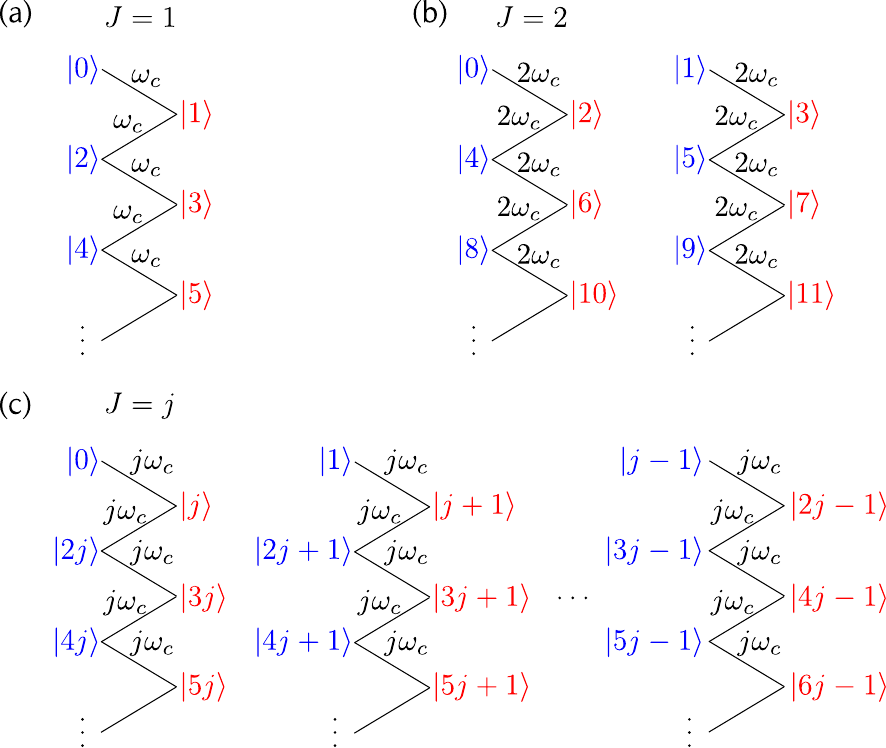}
    \caption{Connection of all the  Fock states $|n\rangle$ and $|m\rangle$ with $g_{n,m}\neq 0$ forms a graph for each given value of $J$. For the $G_J$ given by Eq.~\eqref{eq:G_J}, all the states are \ljy{divided} into two groups $\mathcal{R}$ (in red color) and $\mathcal{B}$ (in blue color) with only states of distinct groups being connected. For the Hamiltonian Eq.~\eqref{eq:H_parameterized} the energy differences between each pair of the connected states are \cc{$J \omega_c$}.}
    \label{fig:bosonic_mode}
\end{figure}

\section{Preparation of multi-squeezed states\label{sec:2}}

Consider a bosonic system initially prepared in one of Fock states, $|n\rangle$, which is an eigenstate of the free bosonic Hamiltonian ($\hbar=1$)
\begin{equation}
    H=\omega_ca^{\dagger}a=\sum_{n=0}^{\infty}n\omega_c|n\rangle\langle n|,
    \label{eq:H_0}
\end{equation}
where $a$ ($a^{\dagger}$) represents the annihilation (creation) operator and $\omega_c$ is the angular frequency of the bosonic mode.

We aim to adiabatically prepare the state $|n_J(\Theta)\rangle$ with
\begin{equation}
|n_J(\lambda)\rangle\equiv e^{-iG_J\lambda}|n\rangle
\label{eq:n_lambda_J}
\end{equation}
by the change of the dimensionless parameter $\lambda\equiv\lambda(t)$ which is a monotonic function with respect to the time $t$. The parameter changes from $\lambda(0)=0$ to $\lambda(T)=\Theta$ for a total evolution time $T$. We use the $J$-th order bosonic interactions
\begin{equation}
    G_J=i\left[\varepsilon \left(a^{\dagger}\right)^J-\varepsilon^*a^J\right],
    \label{eq:G_J}
\end{equation}
where $J$ is a positive integer and $\varepsilon$ is a complex number, for the preparation of multi-squeezed states. 
Note that the matrix elements of $G_{J}$ satisify
\begin{equation}
    g_{n,m}\equiv\langle n|G_J|m\rangle = \langle n_J(\lambda)|i\frac{d}{d\lambda}|m_J(\lambda)\rangle.
    \label{eq:g}
\end{equation}
The Hamiltonian corresponding to multi-squeezed states for adiabatic control reads
\begin{equation}
    H_J(\lambda)=e^{-iG_J\lambda}He^{iG_J\lambda}=\sum_{n=0}^{\infty}E_n|n_J(\lambda)\rangle\langle n_J(\lambda)|,
    \label{eq:H_parameterized}
\end{equation}
where $E_n=n\omega_c$.

When $J=1$, $e^{-iG_1\lambda}$ is a displacement operator, and it gives rise to the first order parameterized Hamiltonian $H_1(\lambda)$ for the bosonic mode, 
\begin{equation}
    \begin{aligned}
            H_1(\lambda)&=e^{-iG_1\lambda}He^{iG_1\lambda}\\
            &=\omega_ca^{\dagger}a-\lambda\omega_c(\varepsilon a^{\dagger}+\varepsilon^*a)+\omega_c|\lambda\varepsilon|^2, 
            \label{eq:H_1}
    \end{aligned}
\end{equation}
where the last term $\omega_c|\lambda\varepsilon|^2$ can be dropped out. 

When $J=2$, $e^{-iG_2\lambda}$ corresponds to a squeezing operator with $\varepsilon=re^{i\vartheta}$ by the use of real parameters $r$ and $\vartheta$. This transforms the original Hamiltonian into
\begin{equation}
    \begin{aligned}
        H_2(\lambda)&=e^{-iG_2\lambda}He^{iG_2\lambda}\\
        &=\omega_c\left[a^{\dagger}a\cosh^2(2\lambda r)+aa^{\dagger}\sinh^2(2\lambda r)\right.\\
        &\left.-\left(a^2e^{-i\vartheta}+\mathrm{H.c.}\right)\sinh(2\lambda r)\cosh(2\lambda r)\right].
        \label{eq:H_2}
    \end{aligned}
\end{equation}

\begin{figure}
    \centering
    \includegraphics[width=8cm]{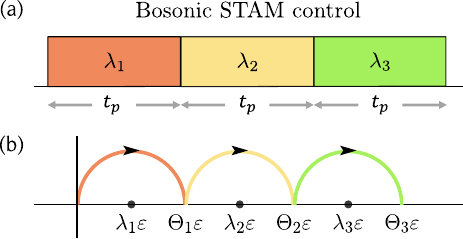}
    \caption{\ljy{STAM method for fast adiabatic control of bosonic states. (a) Control sequence for the bosonic mode. (b) The evolution trajectories of the coherent state in the phase space under the control sequence illustrated in (a).}}
    \label{fig:STAM_coherent}
\end{figure}

In traditional quantum adiabatic control, one slowly varies $\lambda$ from $\lambda=0$ to $\lambda=\Theta$, and in the infinitely slow limit, the initial state $|n_J(0)\rangle=|n\rangle$ will evolve to $|n_J(\Theta)\rangle$. While this adiabatic control has a strong robustness against amplitude fluctuation of the control fields, the long control time limits its application, e.g., due to the short coherence time of the quantum system. 
Here we employ the STAM method~\cite{liu2022shortcuts}, which is originated from the theory of the necessary and sufficient condition of quantum adiabatic evolution~\cite{wang2016necessary}, to prepare the target state $|n_J(\Theta)\rangle$ with unit-fidelity in a fast and robust manner. 

According to Ref.~\cite{wang2016necessary}, the evolution operator $U=\mathcal{T}e^{-i\int_0^{t}H_J dt^\prime}$ (where $\mathcal{T}$ represents time ordering) driven by $H_J$ can be decomposed as
\begin{equation}
    U=U_{\text{adia}}U_{\text{err}},
    \label{eq:U_component}
\end{equation}
where
\begin{equation}
    U_{\text{adia}}(\lambda)=\sum_{n=0}^{\infty}e^{-i\varphi_n}|n_J(\lambda)\rangle\langle n_J(0)|
    \label{eq:U_adia}
\end{equation}
describes the desired ideal adiabatic evolution. 
Since $g_{n,n}=0$ for all $n$ [see Eqs.~\eqref{eq:G_J} and \eqref{eq:g}], each $\varphi_n$ in Eq.~\eqref{eq:U_adia} is the accumulated dynamic phase for the $n$-th state.
Meanwhile, using of the path ordering $\mathcal{P}$ with respect to $\lambda$ in a way similar to the time ordering, the non-adiabatic evolution reads
\begin{equation}
    U_{\text{err}}(\lambda)=\mathcal{P}e^{i\int_0^{\lambda}W(\lambda^\prime)d\lambda^\prime},
    \label{eq:U_err}
\end{equation}
where
\begin{equation}
    W(\lambda) =\sum_{n\neq m}F_{n,m}(\lambda) g_{n,m}|n\rangle\langle m|,
    \label{eq:W}
\end{equation}
with $F_{n,m}(\lambda)=e^{i\left[\varphi_n(\lambda) - \varphi_m(\lambda)\right]}$.

\begin{figure}
    \centering
    \includegraphics[width=8cm]{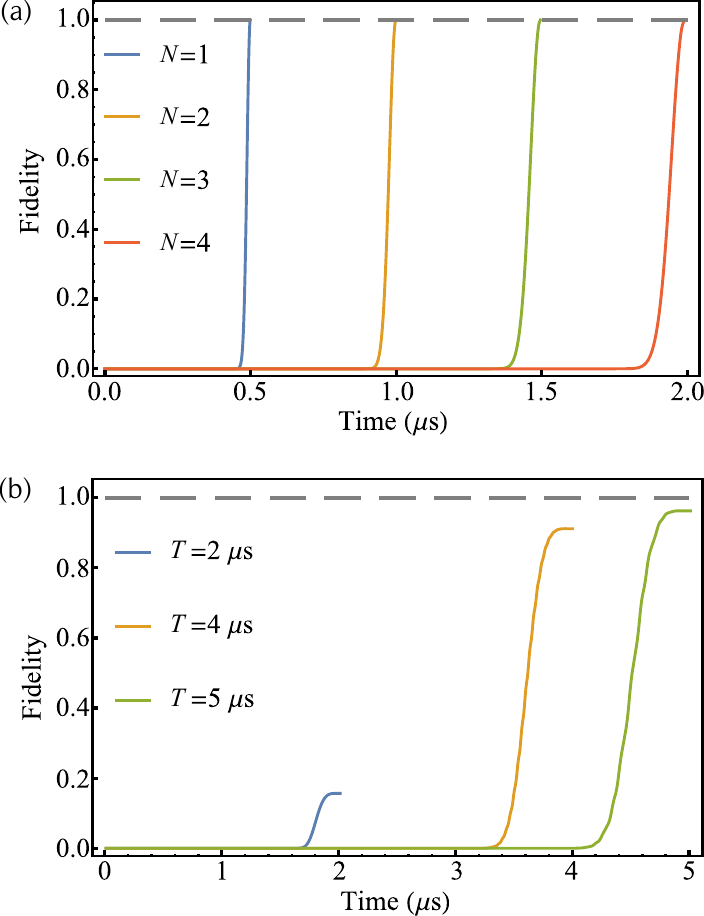}
\caption{Fidelity of the evolved state to the target coherent state during the control \wzy{with $J=1$}. 
(a) The results of our STAM method for different number $N$ of control pulses. 
(b) The results of the traditional adiabatic control for different total evolution time $T$, with
a continuous variation of the control parameter $\lambda(t)=A\exp[-B(t/T-1)^2]-C$, where $A=20.377,B=4,C=0.376974$. Here the parameters $\omega_c=2\pi\times1~\mathrm{MHz}$, $\varepsilon=1$, and $\Theta=20$.}
    \label{fig:Coherent_Preparation_Evolution}
\end{figure}

\begin{figure}
    \centering
    \includegraphics[width=8cm]{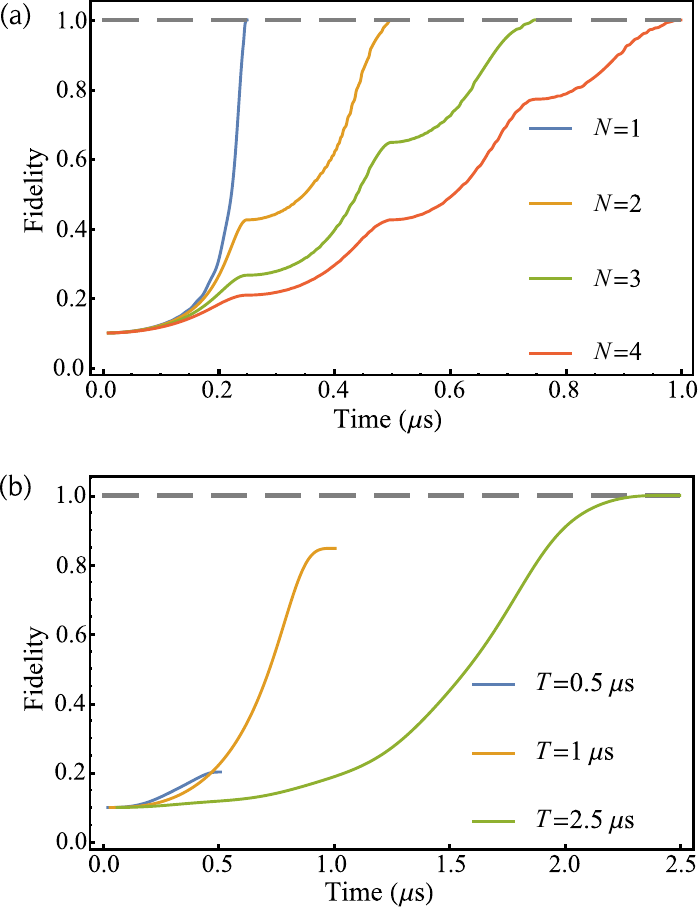}
\caption{Fidelity of the evolved state to the target squeezed state  during the control  \wzy{with $J=2$}. 
(a) The results of our STAM method for different number $N$ of control pulses. 
(b) The results of the traditional adiabatic control for different total evolution time $T$, with
a continuous variation of the control parameter $\lambda(t)=A\exp[-B(t/T-1)^2]-C$, where $A=-0.509165,B=4,C=-0.00916497$. Here $\omega_c=2\pi\times1~\mathrm{MHz}$, $\varepsilon=3i$, and $\Theta=-0.5$.}
    \label{fig:Squeezed_Preparation_Evolution}
\end{figure}

In the following we describe how to choose a functional form of $\lambda(t)$ such that all non-adiabatic effects vanish (i.e., $U_{\text{err}} = I$).

The quantities $g_{n,m}$ with nonzero values form the graphs in Fig.~\ref{fig:bosonic_mode}, where the Fock states $|n\rangle$ and $|m\rangle$ are either connected ($g_{n,m}\neq0$) or disconnected ($g_{n,m}=0$). In our case, only
\begin{equation}
    g_{n+J,n}=i\varepsilon\sqrt{\frac{(n+J)!}{n!}}
    \label{eq:g_n+J_n}
\end{equation}
and $g_{n,n+J}=g_{n+J,n}^*$ with $n\geqslant 0$ have nonzero values.
An important property of these graphs is that all the Fock states can be \wzy{divided} into two groups: group $\mathcal{R}$ and group $\mathcal{B}$, illustrated in red and blue colors in Fig.~\ref{fig:bosonic_mode}, respectively. Additionally, only states from different groups can be connected. Therefore, if we keep the dynamic phases $\varphi_n(\lambda)=2k_n(\lambda)\pi$ for $n\in \mathcal{B}$ while $\varphi_m(\lambda)=k_m(\lambda)\pi+\pi$ for $m\in \mathcal{R}$ with both $k_n(\lambda)$  and $k_m(\lambda)$ being integers along the path, we can write
\begin{equation}
    W(\lambda) =F(\lambda) \sum_{n\neq m} g_{n,m}|n\rangle\langle m|,
    \label{eq:W_rewrite}
\end{equation}
with a common  $F(\lambda)\in \{+1,-1\}$. In this manner, $W(\lambda)$ at different $\lambda$ commute \wzy{and hence $U_{\text{err}}(\lambda)=e^{i\int_0^{\lambda}W(\lambda^\prime)d\lambda^\prime}$, where the path ordering has been removed. All non-adiabatic errors are eliminated at the end of the evolution time $T$, i.e., $U_{\text{err}}(\Theta)=I$} when
\begin{equation}
    \int_0^{\Theta} F(\lambda)d\lambda=0.
    \label{eq:F0}
\end{equation}
We achieve these by applying $H_J(\lambda)$ [Eq.~\eqref{eq:H_parameterized}] at  $N$ equally-spaced points of the parameter $\lambda=\lambda_k$ subsequently  with $k=1,2,\ldots,N$. Here
\begin{eqnarray}
    \lambda_k=\frac{\Theta}{2N}(2k-1).
    \label{eq:lambda_k}
\end{eqnarray}
The control at each $\lambda_k$ has a time duration $t_p=\pi/(J\omega_c)$ such that each bosonic control pulse
\begin{equation}
    P_k\equiv e^{-i H_J(\lambda_k) t_p} \label{eq:propagator}
\end{equation}
introduces a sign change to $F(\lambda)$ at $\lambda=\lambda_k$, because the directly connected states in Fig.~\ref{fig:bosonic_mode} have an energy difference of $J\omega_c$. \wzy{That is, $F(\lambda)=(-1)^k$ for $\lambda\in[\lambda_k,\lambda_{k+1})$ with $\lambda_0\equiv 0$ and $\lambda_{N+1}\equiv\Theta$. Since the sequence Eq.~\eqref{eq:lambda_k} satisfies Eq.~\eqref{eq:F0}, we have} $U_{\text{err}}(\Theta) = I$ at the end of the evolution, i.e.,
\begin{equation}
    U(\Theta) = P_N P_{N-1} \cdots P_2 P_1 = U_{\text{adia}}(\Theta). \label{eq:U_seq}
\end{equation}
Because  $\int_0^{\Theta_k} F(\lambda)d\lambda=0$ for any
\begin{eqnarray}
    \Theta_k\equiv k\Theta/N, ~(k=0,1,2,\ldots,N)
    \label{eq:Theta_k}
\end{eqnarray}
we also achieve the target adiabatic evolution perfectly without any non-adiabatic errors (i.e., $U=U_{\text{adia}}$) at all these parameter points $\lambda = \Theta_k$ ($k=1,2,\ldots,N$). 

\begin{figure}
    \centering
    \includegraphics[width=8cm]{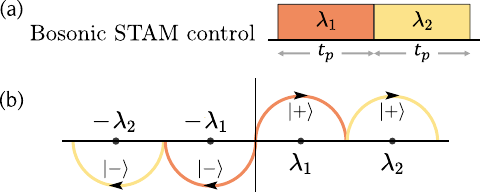}
    \caption{\cc{Preparation of hybrid entangled state by STAM control on a bosonic mode coupled with a qubit. (a) Control sequence for bosonic STAM. (b) Evolution of the coherent state under the control sequence illustrated in (a) is conditional to the qubit states $|+\rangle$ and  $|-\rangle$.}}
    \label{fig:STAM_evolution}
\end{figure}

\cc{The bosonic control pulse $P_k$ [Eq.~\eqref{eq:propagator}] for the Gaussian cases (i.e., $J=1,2$) can be implemented in various systems~\cite{anton2019ultrastrong,andersen2016squeezing}.
For the non-Gaussian cases where $J\geqslant3$, the realization of the control Hamiltonian $H_J(\lambda)$ is more challenging. A way to implement $P_k$ is to use the composite pulse
\begin{eqnarray}
    P_k=e^{-iG_J\lambda_k}e^{-i\omega_c a^{\dagger}a t_p}e^{iG_J\lambda_k},
    \label{eq:P_Decompose}
\end{eqnarray}
where the transformations $e^{-iG_J\lambda_k}$ and $e^{iG_J\lambda_k}$ can be realized by the protocol in Ref.~\cite{mcconnell2022multisqueezed}, e.g., in strongly coupled superconducting qubits and microwave resonators~\cite{mei2013analog,stojanovic2019quantum,nauth2023spectral,casanova2018connecting}.
}

\begin{figure}
    \centering
    \includegraphics[width=8cm]{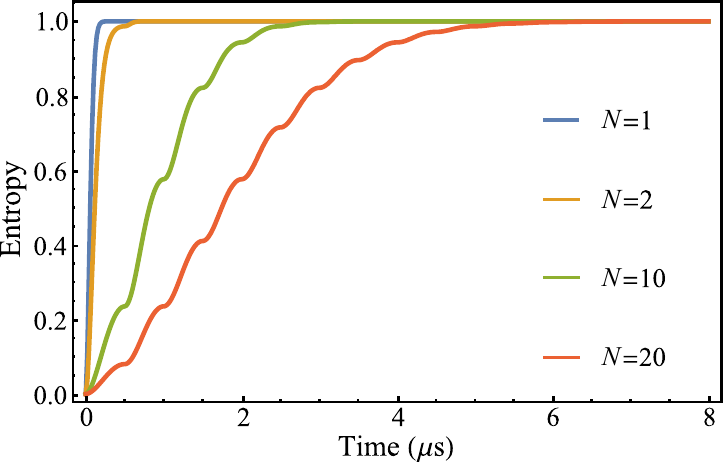}
    \caption{Entanglement entropy evolution of hybrid entangled state using the STAM method illustrated in Fig.~\ref{fig:STAM_evolution}, \wzy{for different pulse number $N$, $\Theta=2$, and $\omega_c=2\pi\times1~\mathrm{MHz}$.}}
    \label{fig:entropy_evolution}
\end{figure}

%

Our method avoids the long evolution time required by the traditional adiabatic methods that use slowly varying Hamiltonians. We can generate a $J$-th order bosonic quantum state from the vacuum state $|0\rangle$ or other initial states, by applying the control sequence  Eq.~\eqref{eq:U_seq}.

\wzy{A physical picture of how the multi-squeezed state is prepared by our method is provided in  Fig.~\ref{fig:STAM_coherent} by considering the case of $J=1$ and the initial state being the vacuum state $|0\rangle$ (which is a coherent state).  As illustrated in Fig.~\ref{fig:STAM_coherent}, during each control pulse $P_{k}$ the state (which remains a coherent state) rotates clockwise around the point $\lambda_k\varepsilon$ with the radius $|(\Theta_{k-1}-\lambda_k)\varepsilon|$ and an angular frequency $\omega_c$ in the phase space as~\cite{yang2011preserving} 
\begin{eqnarray}
    e^{-iH_1(\lambda_k)t}|\Theta_{k-1}\varepsilon\rangle=|\Phi(t)\varepsilon\rangle e^{-i\theta},
    \label{eq:alpha_evolution}
\end{eqnarray}
where $\Phi(t)=\left(\Theta_{k-1}-\lambda_k\right)e^{-i\omega_ct}+\lambda_k$ and the global phase $\theta$ can be neglected. When $t=t_p$, the pulse $P_k$ is complected and it transforms the coherent state $|\Theta_{k-1}\varepsilon\rangle$, which corresponds to the ground state of $H_1(\Theta_{k-1})$, into the ground state of $H_1(\Theta_{k})$.  }

\wzy{To demonstrate that our method is much faster than traditional techniques for achieving the same outcome, in} Figs.~\ref{fig:Coherent_Preparation_Evolution} and \ref{fig:Squeezed_Preparation_Evolution}, we compare our STAM method with the traditional adiabatic control which uses a continuous variation of parameters, \wzy{for $J=1$ and $J=2$, respectively. The results were simulated by using} QuTiP~\cite{johansson2012QuTiP2,johansson2012QuTiP3}. Figure~\ref{fig:Coherent_Preparation_Evolution} shows the fidelity of coherent-state preparation during the control for $|\wzy{\Theta\varepsilon}=20\rangle$. Our STAM method provides a much faster and much higher fidelity than the traditional adiabatic control. In Fig.~\ref{fig:Coherent_Preparation_Evolution}, a fidelity of $100\%$ can be reached within $0.5$ $\mu$s by using our method, while for the traditional adiabatic control, the time for a fidelity higher than $95\%$ is approximately $5$ $\mu$s, \wzy{an order of magnitude longer than} the time used in our method. Note that our method performs much better than the traditional adiabatic control when the amplitude $\Theta$ is larger, since the time required in our method is a fixed pulse duration $t_p$ while for the traditional adiabatic control needs more time for a larger $\Theta$. \wzy{S}imilarly, Fig.~\ref{fig:Squeezed_Preparation_Evolution} demonstrates the advantages of our STAM method for the preparation of squeezed vacuum state $|\cc{n=}0,\xi=3i\rangle$. These results indicate that the STAM method can achieve the target states in a much shorter time than the traditional adiabatic control.

The idea of our method can be generalized to other quantum systems. In the following section we demonstrate the preparation of entangled states for a hybrid quantum system consisting of a qubit and a bosonic mode using our method. 

\section{Preparation of hybrid entangled states\label{sec:3}}

In this section, we consider a hybrid system consisting of a bosonic mode and a qubit \cc{with the Hamiltonian
\begin{equation}
    \begin{aligned}
        H^{\prime}_1(\lambda)=\omega_c a^{\dagger}a-\lambda\omega_c(a^{\dagger}+a)\sigma_x,
        \label{eq:H_prime}
    \end{aligned}
\end{equation}
where $\sigma_x=|e\rangle\langle g|+|g\rangle\langle e|$ with $\left\{|e\rangle,\,|g\rangle\right\}$ being the qubit states. 
This type of interaction could be realized across various platforms, including superconducting circuit~\cite{anton2019ultrastrong,vlastakis2013deterministically,bild2022schrodinger,yoshihara2017superconducting}, trapped atoms~\cite{monroe1996schrodinger,johnson2017ultrafast,hacker2019deterministic,koch2023QRM}, and neutral atoms~\cite{plodzien2018rydberg,stojanovic2021scalable}.}

Our goal is to prepare a hybrid entangled state of the form
\begin{eqnarray}
    |\psi\rangle_{AB}=\frac{1}{\sqrt{2}}(|\alpha\rangle_A|+\rangle_B\pm|-\alpha\rangle_A|-\rangle_B),
    \label{eq:hybrid_entangled_state}
\end{eqnarray}
where $|\pm\alpha\rangle$ represent the coherent states of the bosonic mode and $|\pm\rangle=\frac{1}{\sqrt{2}}(|e\rangle\pm|g\rangle)$ are the eigenstates of the Pauli matrix \cc{$\sigma_x$.
This pure state} is known as the Schrödinger-cat state~\cite{monroe1996schrodinger,wineland2013nobel,
vlastakis2013deterministically,johnson2017ultrafast,
sychev2017enlargement,bild2022schrodinger,
liao2016generation,hacker2019deterministic,yoshihara2017superconducting}. 

\ljy{\subsection{STAM without qubit control}}

%

\begin{figure}
    \centering
    \includegraphics[width=8.65cm]{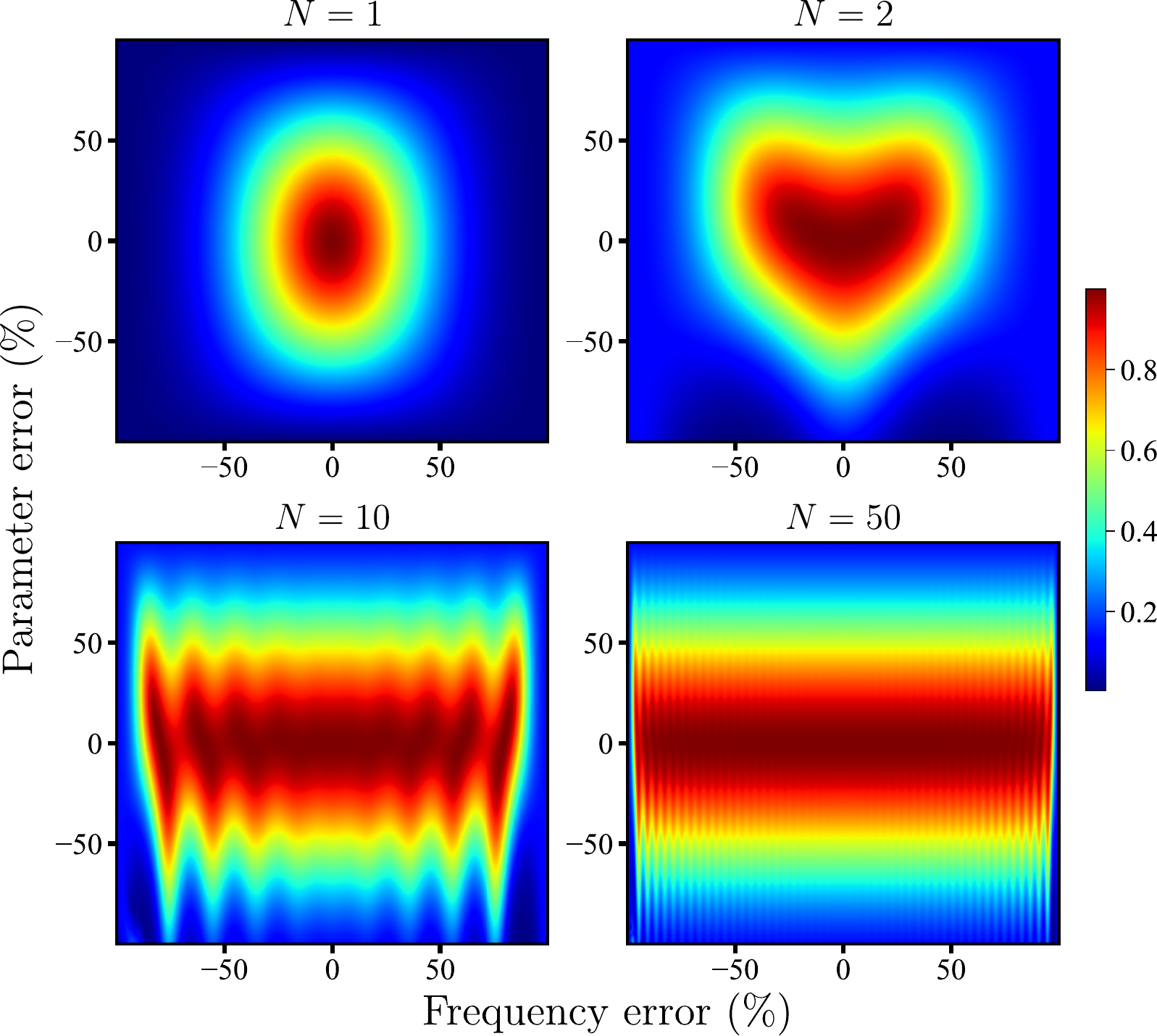}
    \caption{\cc{Fidelity of hybrid entangled state preparation via the STAM control illustrated in Fig.~\ref{fig:STAM_evolution} for various numbers $N$ of bosonic control pulses when there are static relative fluctuations to the ideal sequence parameters $\lambda_k$ [Eq.~\eqref{eq:lambda_k}] and the ideal frequency $\omega_c=2\pi \times 100$ MHz. Here $\Theta=2$.}}
    \label{fig:STAM_error}
\end{figure}

\begin{figure}
    \centering
    \includegraphics[width=8cm]{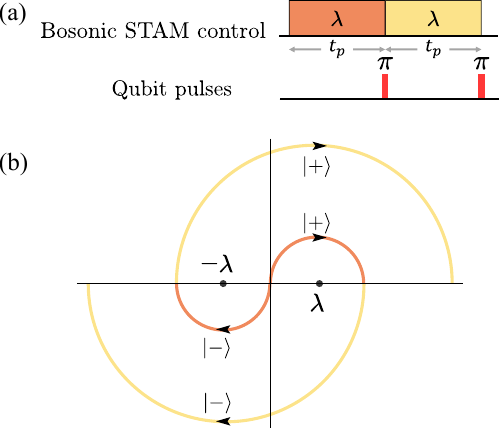}
    \caption{\cc{Large hybrid entangled state realized by synchronized STAM control and qubit control. (a) Bosonic STAM and qubit control sequence. (b) The control sequence illustrated in (a) amplifies the trajectories of the coherent state evolution.}}
    \label{fig:flip_evolution}
\end{figure}

Regarding to the states $|\pm\rangle$ of the qubit, the bosonic Hamiltonian \wzy{Eq.~\eqref{eq:H_prime} becomes}
\begin{eqnarray}
    H_\pm(\lambda)=\omega_c a^{\dagger}a\mp\lambda\omega_c (a^{\dagger}+a),
    \label{eq:Ham_pm}
\end{eqnarray}
which corresponds to Eq.~\eqref{eq:H_1} with $\varepsilon=1$. 

\ljy{T}herefore, according to the theory in Sec.~\ref{sec:2}, applying the Hamiltonian $H^{\prime}_1(\lambda_k)$
 in Eq.~\eqref{eq:H_prime} for a time duration $t_p=\pi/\omega_c$ would generate the pulse
\begin{equation}
    \begin{aligned}
        P_k^{\prime}= e^{-i H_+(\lambda_k) t_p } |+\rangle\langle+|+ e^{-i H_-(\lambda_k) t_p }|-\rangle\langle-|,
        \label{eq:P_k}
    \end{aligned}
\end{equation}
which transforms the given initial state $|0\rangle|g\rangle$ to a hybrid entangled state
\begin{equation}
    \begin{aligned}
        |\psi\rangle=\frac{1}{\sqrt{2}}\left(|\Theta_k\rangle|+\rangle-|-\Theta_k\rangle|-\rangle\right).
        \label{eq:psi}
    \end{aligned}
\end{equation}
\wzy{For the STAM illustrated in Fig.~\ref{fig:STAM_evolution}, we achieve} the target hybrid entangled state
\begin{equation}
        |\psi_{\text{target}}\rangle=\frac{1}{\sqrt{2}}\left(|\Theta\rangle|+\rangle-|-\Theta\rangle|-\rangle\right).
        \label{eq:psi_target}
\end{equation}

\wzy{We first consider the evolution of the von Neumann entanglement entropy~\cite{vedral1997quantifying,amico2008entanglement,bengtsson2017geometry}
between the bosonic mode and the qubit during the STAM control. The von Neumann entanglement entropy 
is calculated as $\mathcal{S}(\rho_A)=-\operatorname{Tr}\left[\rho_A\log_2\left(\rho_A\right)\right]$, where $\rho_A=\operatorname{Tr}_B\left(\rho_{AB}\right)$ and $\rho_B=\operatorname{Tr}_A\left(\rho_{AB}\right)$ represent the reduced density matrices for each partition~\cite{vedral1997quantifying,amico2008entanglement,bengtsson2017geometry}. As Eq.~\eqref{eq:hybrid_entangled_state}, here we refer the bosonic mode (qubit) as the subsystem $A$ ($B$). For the given final state $|\psi_{\text{target}}\rangle$ with $\Theta=2$ in Eq.~\eqref{eq:psi_target}, Fig.~\ref{fig:entropy_evolution} shows the evolution of the von Neumann entanglement entropy. The entanglement between the bosonic mode and the qubit can reach its maximum in a short time.}

In Fig.~\ref{fig:STAM_error} we \wzy{further} demonstrate the fidelity between the target state, Eq.~\eqref{eq:psi_target}, and the prepared state using the STAM sequence in Fig.~\ref{fig:STAM_evolution}. The simulated results show that our STAM sequence gives a high fidelity of the state preparation. \wzy{The robustness of the sequence against pulse errors is better for a larger number $N$ of pulses.}

\ljy{\subsection{STAM with qubit control}}

\begin{figure}
    \centering
    \includegraphics[width=8cm]{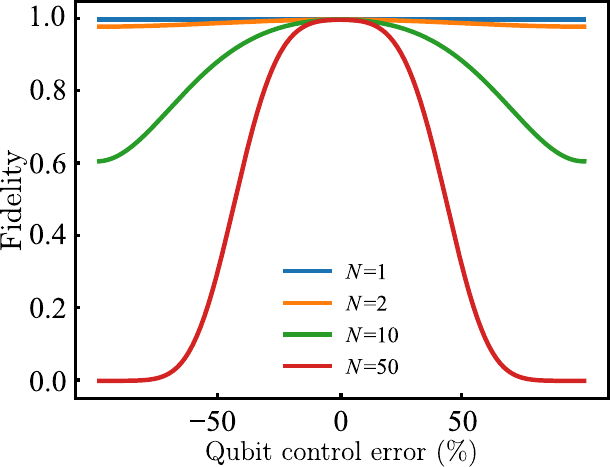}
    \caption{\cc{Fidelity of hybrid entangled state preparation via the STAM with qubit control illustrated in Fig.~\ref{fig:flip_evolution} as a function of control error $\delta$ for various numbers $N$ of bosonic pulses. The control amplitude $(1+\delta)\Omega$ of the qubit control has a static fluctuation with respect to the ideal one $\Omega=2\pi\times50$ MHz. Here $\omega_c=2\pi\times100~\mathrm{kHz}$ and $\lambda=0.05$.}}
    \label{fig:flip_error}
\end{figure}

We note that the STAM sequence in Fig.~\ref{fig:STAM_evolution} restricts the possibility of preparing coherent states with an arbitrary large $\Theta$,  because of the limited values of $\lambda$ in the Hamiltonian. \wzy{To address this challenge, we apply a $\pi$ pulse on the qubit after
each STAM pulse on the bosonic mode, see Fig.~\ref{fig:flip_evolution} (a), following a similar idea in Ref.~\cite{monroe1996schrodinger}. Each qubit $\pi$ pulse swaps the qubit states $|+\rangle$ and $|-\rangle$ and hence realizes the transformation $\sigma_x \rightarrow -\sigma_x$. After the application of $k$ qubit $\pi$ pulses, the Hamiltonian Eq.~\eqref{eq:H_prime} becomes
\begin{eqnarray}
    H_I=\omega_ca^{\dagger}a-(-1)^{k}\lambda\omega_c(a^{\dagger}+a)\sigma_x.
\end{eqnarray}
As illustrated in Fig.~\ref{fig:flip_evolution}, each qubit $\pi$ pulse swaps the qubit states $|+\rangle$ and $|-\rangle$ and hence the rotating centers of the conditional bosonic evolution.
Consequently, the achieved value $\Theta=2N\lambda$ of the target state [Eq.~\eqref{eq:psi_target}] can be arbitrarily large by increasing the number of bosonic pulses $N$ while keeping parameter fixed, i.e., $\lambda_k\equiv\lambda$ [see Fig.~\ref{fig:flip_evolution}].
To enhance the robustness of the control on the qubit, we apply the qubit control Hamiltonian $H_{\text{ctrl}}=(-1)^{k}\frac{\Omega}{2}\sigma_z$ for the $k$-th qubit $\pi$ pulse, where  
the alternation of the sign mitigates potential amplitude fluctuation of the control on the qubit.
}

Figure~\ref{fig:flip_error} illustrates the simulated fidelity between the prepared state with the target states $|\psi_{\text{target}}\rangle$ [Eq.~\eqref{eq:psi_target}] using the sequence in Fig.~\ref{fig:flip_evolution}. The results show that one can reliably prepare the target state $|\psi_{\text{target}}\rangle$ with increasing values of $\Theta=2N\lambda$. 
The state fidelity remains very high even with a relatively large static error \wzy{in the qubit control $\Omega \rightarrow (1+\delta)\Omega$}. Even when $N=50$ and \wzy{the} error reaches $\delta=\pm25\%$, the fidelity still remains above $90\%$.

\section{Conclusion and outlook\label{sec:conclusion}}

In summary, we have proposed a STAM method to greatly speed up the adiabatic preparation of the multi-squeezed states of a bosonic mode as well as the entanglement of these states with a qubit. Our STAM method achieves the adiabatic speedup
by dynamically adjusting the Hamiltonian parameters such that the non-adiabatic effects are coherently eliminated. Using our method, the target multi-squeezed states can be prepared in a short finite time, e.g., less than $10\%$ of the time required in the traditional adiabatic control as demonstrated by our numerical simulations. For the case of Schrödinger-cat state preparation in a hybrid system of a qubit and a bosonic mode, we show linear increase of the amplitude of the Schrödinger-cat state with the number of pulses. This enables fast preparation of arbitrarily high amplitude entangled multi-squeezed states. The STAM method has some intrinsic robustness against some kinds of errors. \cc{The robustness of our method could be further enhanced by numerical optimization methods such as the gradient ascent pulse engineering (GRAPE) algorithm~\cite{khaneja2005optimal,wu2015optimal} and other advanced techniques such as 
enhanced shortcuts to adiabaticity (eSTA) methods as discussed in recent research~\cite{whitty2020eSTA,whitty2022robust,sascha2022coherent,manuel2023spin}.}

\begin{acknowledgments}
This work was supported by National Natural Science Foundation of China (Grant No.~12074131), the Natural Science Foundation of Guangdong Province (Grant No.~2021A1515012030), and National College Student Innovation and Entrepreneurship Training Program (Grant No.~202310574055). C.C. and J.-Y.L. contributed equally to this work.
\end{acknowledgments}


\begin{thebibliography}{99}
\bibitem{messiah1961book} A. Messiah, \textit{Quantum Mechanics, Volume II} (North-Holland, Amsterdam, 1961).
\bibitem{berry1984quantal} M. V. Berry, Quantal phase factors accompanying adiabatic changes, \href{http://doi.org/10.1098/rspa.1984.0023}{Proc. R. Soc. A \textbf{392}, 45 (1984)}.
\bibitem{albash2018adiabatic} T. Albash and D. A. Lidar, Adiabatic quantum computation, \href{https://doi.org/10.1103/RevModPhys.90.015002}{Rev. Mod. Phys.  \textbf{90}, 015002 (2018)}.
\bibitem{zheng2012squeezing} S.-B. Zheng, Generation of atomic and field squeezing by adiabatic passage and symmetry breaking, \href{https://link.aps.org/doi/10.1103/PhysRevA.86.013828}{Phys. Rev. A \textbf{86}, 013828 (2012)}.
\bibitem{leroux2018enhancing} C. Leroux, L. C. G. Govia, and A. A. Clerk, Enhancing cavity quantum electrodynamics via antisqueezing: Synthetic ultrastrong coupling, \href{https://link.aps.org/doi/10.1103/PhysRevLett.120.093602}{Phys. Rev. Lett. \textbf{120}, 093602 (2018)}.
\bibitem{chen2019squeezing} J. Chen, D. Konstantinov, and K. Mølmer, Adiabatic preparation of squeezed states of oscillators and large spin systems coupled to a two-level system, \href{https://link.aps.org/doi/10.1103/PhysRevA.99.013803}{Phys. Rev. A \textbf{99}, 013803 (2019)}.
\bibitem{andersen2016squeezed} U. L. Andersen, T. Gerhring, C. Marquardt, and G. Leuchs, 30 years of squeezed light generation, \href{https://doi.org/10.1088/0031-8949/91/5/053001}{Phys. Scr. \textbf{91}, 053001 (2016)}.
\bibitem{demirplak2003adiabatic} M. Demirplak and S. A. Rice, Adiabatic population transfer with control fields, \href{https://doi.org/10.1021/jp030708a} {J. Phys. Chem. A \textbf{107}, 9937 (2003)}.
\bibitem{berry2009transitionless}  M. V. Berry, Transitionless quantum driving, \href{https://doi.org/10.1088/1751-8113/42/36/365303}{J. Phys. A: Math. Theor. \textbf{42}, 365303 (2009)}. 
\bibitem{chen2010shortcut}  X. Chen, I. Lizuain, A. Ruschhaupt, D. Guéry-Odelin, and J. G. Muga, Shortcut to adiabatic passage in two- and three-level atoms, 
\href{https://link.aps.org/doi/10.1103/PhysRevLett.105.123003}{Phys. Rev. Lett. \textbf{105}, 123003 (2010)}.
\bibitem{torrontegui2013sta} E. Torrontegui, S. Ib{\'a}{\~n}ez, S. {Mart{\'i}nez-Garaot}, M. Modugno, A. del Campo, D. {Gu{\'e}ry-Odelin}, A. Ruschhaupt, X. Chen, and J. G. Muga, Shortcuts to adiabaticity, \href{https://doi.org/10.1016/B978-0-12-408090-4.00002-5}{Adv. At. Mol. Opt. Phys. \textbf{62}, 117 (2013)}. 
\bibitem{odelin2019shortcuts} D. Guéry-Odelin, A. Ruschhaupt, A. Kiely, E. Torrontegui, S. Martínez-Garaot, and J. G. Muga, Shortcuts to adiabaticity: Concepts, methods, and applications, \href{https://link.aps.org/doi/10.1103/RevModPhys.91.045001}{Rev. Mod. Phys. \textbf{91}, 045001 (2019)}.
\bibitem{bason2012high} M. G. Bason, M. Viteau, N. Malossi, P. Huillery, E. Arimondo, D. Ciampini, R. Fazio, V. Giovannetti, R. Mannella, and O. Morsch, High-fidelity quantum driving, \href{https://doi.org/10.1038/nphys2170}{Nat. Phys. \textbf{8}, 147 (2012)}.
\bibitem{baksic2016sta} A. Baksic, H. Ribeiro, and A. A. Clerk, Speeding up adiabatic quantum state transfer by using dressed states, \href{https://link.aps.org/doi/10.1103/PhysRevLett.116.230503}{Phys. Rev. Lett. \textbf{116}, 230503 (2016)}.
\bibitem{zhou2017accelerated} B. B. Zhou, A. Baksic, H. Ribeiro, C. G. Yale, F. J. Heremans, P. C. Jerger, A. Auer, G. Burkard, A. A. Clerk, and D. D. Awschalom, Accelerated quantum control using superadiabatic dynamics in a solid-state lambda system, \href{https://doi.org/10.1038/nphys3967}{Nat. Phys. \textbf{13}, 330 (2016)}.
\bibitem{du2016experimental}Y.-X. Du, Z.-T. Liang, Y.-C. Li, X.-X. Yue, Q.-X. Lv, W. Huang, X. Chen, H. Yan, and S.-L. Zhu, Experimental realization of stimulated Raman shortcut-to adiabatic passage with cold atoms, \href{https://doi.org/10.1038/ncomms12479}{Nat. Commun. \textbf{7}, 12479 (2016)}.
\bibitem{song2021robust} X.-K. Song, F. Meng, B.-J. Liu, D. Wang, L. Ye, and M.-H. Yung, Robust stimulated Raman shortcut-to-adiabatic passage with invariant-based optimal control, \href{https://doi.org/10.1364/OE.417343}{Opt. Express \textbf{29}, 7998 (2021)}.
\bibitem{liu2022shortcuts}  Y. Liu and Z.-Y. Wang, Shortcuts to adiabaticity with inherent robustness and without auxiliary control, \href{https://doi.org/10.48550/arXiv.2211.02543}{arXiv:2211.02543}.
\bibitem{wang2016necessary}  Z.-Y. Wang and M. B. Plenio, Necessary and suﬀicient condition for quantum adiabatic evolution by unitary control fields, \href{https://link.aps.org/doi/10.1103/PhysRevA.93.052107}{Phys. Rev. A \textbf{93}, 052107 (2016)}.
\bibitem{xu2019geodesics} K. Xu, T. Xie, F. Shi, Z.-Y. Wang, X. Xu, P. Wang, Y. Wang, M. B. Plenio, and J. Du, Breaking the quantum adiabatic speed limit by jumping along geodesics, \href{https://www.science.org/doi/10.1126/sciadv.aax3800}{Sci. Adv. \textbf{5}, eaax3800 (2019)}.
\bibitem{zheng2022accelerated} W. Zheng, J. Xu, Z. Wang, Y. Dong, D. Lan, X. Tan, and Y. Yu, Accelerated Quantum Adiabatic Transfer in Superconducting Qubits, \href{https://doi.org/10.1103/PhysRevApplied.18.044014}{
Phys. Rev. Applied \textbf{18}, 044014 (2022)}.
\bibitem{gong2023jumping} M. Gong, M. Yu, R. Betzholz, Y. Chu, P. Yang, Z. Wang, and J. Cai, Accelerated quantum control in a three-level system by jumping along the geodesics, \href{https://link.aps.org/doi/10.1103/PhysRevA.107.L040602}{Phys. Rev. A \textbf{107}, L040602 (2023)}.
\bibitem{xu2024control} S. Xu, C. Xie, and Z.-Y. Wang, Enhancing electron-nuclear resonances by dynamical control switching, \href{https://link.aps.org/doi/10.1103/PhysRevA.109.L020601}{Phys. Rev. A \textbf{109}, L020601 (2024)}.
\wzy{\bibitem{zeng2024wide}  K. Zeng, X. Yu, M. B. Plenio, and Z.-Y. Wang, Wide-band Unambiguous Quantum Sensing via Geodesic Evolution, \href{https://doi.org/10.48550/arXiv.2307.10537}{arXiv:2307.10537}.}
\bibitem{fisher1984impossibility}  R. A. Fisher, M. M. Nieto, and V. D. Sandberg, Impossibility of naively generalizing squeezed coherent states, \href{https://link.aps.org/doi/10.1103/PhysRevD.29.1107}{Phys. Rev. D \textbf{29}, 1107 (1984)}.
\bibitem{hillery1984nonlinear} M. Hillery, M. S. Zubairy, and K. Wódkiewicz, Squeezing in higher order nonlinear optical processes, \href{https://doi.org/10.1016/0375-9601(84)90120-8}{Phys. Lett. A \textbf{103}, 259 (1984)}.
\bibitem{braunstein1987squeezing}  S. L. Braunstein and R. I. McLachlan, Generalized squeezing, \href{https://link.aps.org/doi/10.1103/PhysRevA.35.1659}{Phys. Rev. A \textbf{35}, 1659 (1987)}.
\bibitem{hillery1990divergence} M. Hillery, Photon number divergence in the quantum theory of n-photon down conversion, \href{https://link.aps.org/doi/10.1103/PhysRevA.42.498}{Phys. Rev. A \textbf{42}, 498 (1990)}.
\bibitem{vraunstein1990statistics}  S. L. Braunstein and C. M. Caves, Phase and homodyne statistics of generalized squeezed states, \href{https://link.aps.org/doi/10.1103/PhysRevA.42.4115}{Phys. Rev. A \textbf{42}, 4115 (1990)}.
\bibitem{vourdas1992squeezing} A. Vourdas, Generalized squeezing, Bogoliubov quasiparticles, and information in two- and three-mode systems, \href{https://link.aps.org/doi/10.1103/PhysRevA.46.442}{Phys. Rev. A \textbf{46}, 442 (1992)}.
\bibitem{walls1963squeezed}  D. F. Walls, Squeezed states of light, \href{https://doi.org/10.1038/306141a0}{Nature(London) \textbf{306}, 141 (1983)}.
\bibitem{glauber1963coherent}  R. J. Glauber, Coherent and incoherent states of the radiation field, \href{https://link.aps.org/doi/10.1103/PhysRev.131.2766}{Phys. Rev. \textbf{131}, 2766 (1963)}.
\bibitem{zhang1990coherentReview}  W.-M. Zhang, D. H. Feng, and R. Gilmore, Coherent states: Theory and some applications, \href{https://link.aps.org/doi/10.1103/RevModPhys.62.867}{Rev. Mod. Phys. \textbf{62}, 867 (1990)}.
\bibitem{chang2020experiment} C. W. S. Chang, C. Sabín, P. Forn-Díaz, F. Quijandría, A. M. Vadiraj, I. Nsanzineza, G. Johansson, and C. M. Wilson, Observation of three-photon spontaneous parametric down-conversion in a superconducting parametric cavity, \href{https://link.aps.org/doi/10.1103/PhysRevX.10.011011}{Phys. Rev. X \textbf{10}, 011011 (2020)}.
\bibitem{banaszek1997interference}  K. Banaszek and P. L. Knight, Quantum interference in three-photon down-conversion, \href{https://link.aps.org/doi/10.1103/PhysRevA.55.2368}{Phys. Rev. A \textbf{55}, 2368 (1997)}.
\bibitem{albarelli2018wigner} F. Albarelli, M. G. Genoni, M. G. A. Paris, and A. Ferraro, Resource theory of quantum non-gaussianity and wigner negativity, \href{https://link.aps.org/doi/10.1103/PhysRevA.98.052350}{Phys. Rev. A \textbf{98}, 052350 (2018)}.
\bibitem{zheng2021conversion}  Y. Zheng, O. Hahn, P. Stadler, P. Holmvall, F. Quijandría, A. Ferraro, and G. Ferrini, Gaussian conversion protocols for cubic phase state generation, \href{https://link.aps.org/doi/10.1103/PRXQuantum.2.010327}{PRX Quantum \textbf{2}, 010327 (2021)}.
\bibitem{mcconnell2022multisqueezed} P. McConnell, A. Ferraro, and R. Puebla, Multi-squeezed state generation and universal bosonic control via a driven quantum Rabi model, \href{https://doi.org/10.48550/arXiv.2209.07958}{arXiv:2209.07958}.
\ljy{\bibitem{wenlong2021circuit} W. Ma, S. Puri, R. J. Schoelkopf, M. H. Devoret, S.M. Girvin, and L. Jiang, Quantum control of bosonic modes with superconducting circuits, \href{https://doi.org/10.1016/j.scib.2021.05.024}{Sci. Bull. \textbf{66}, 1789 (2021)}.
\bibitem{anton2019ultrastrong} A. Frisk Kockum, A. Miranowicz, S. De Liberato, S. Savasta, and F. Nori, Ultrastrong coupling between light and matter, \href{https://doi.org/10.1038/s42254-018-0006-2}{Nat. Rev. Phys. \textbf{1}, 19-40 (2019)}.
\bibitem{andersen2016squeezing} U. L. Andersen, T. Gehring, C. Marquardt, and G. Leuchs, 30 years of squeezed light generation, \href{https://doi.org/10.1088/0031-8949/91/5/053001}{Phys. Scr. \textbf{91}, 053001 (2016)}.}
\cc{\bibitem{mei2013analog} F. Mei, V. M. Stojanović, I. Siddiqi, and L. Tian, Quantum dynamics of the small-polaron formation in a superconducting analog simulator, \href{http://dx.doi.org/10.1103/PhysRevB.88.224502}{Phys. Rev. B \textbf{88}, 224502 (2013)}.
\bibitem{stojanovic2019quantum} V. M. Stojanović and I. Salom, Quantum dynamics of the small-polaron formation in a superconducting analog simulator, \href{https://doi.org/10.1103/PhysRevB.99.134308}{Phys. Rev. B \textbf{99}, 134308 (2019)}.
\bibitem{nauth2023spectral} J. K. Nauth and V. M. Stojanović, Quantum dynamics of the small-polaron formation in a superconducting analog simulator, \href{https://doi.org/10.1103/PhysRevB.107.174306}{Phys. Rev. B \textbf{107}, 174306 (2023)}.}
\ljy{\bibitem{casanova2018connecting} J. Casanova, R. Puebla, H. Moya-Cessa, M. B. Plenio, Connecting nth order generalised quantum Rabi models: Emergence of nonlinear spin-boson coupling via spin rotations, \href{https://doi.org/10.1038/s41534-018-0096-9}{npj Quantum Inf. \textbf{4}, 47 (2018)}.}
\bibitem{yang2011preserving} W. Yang, Z.-Y. Wang, and R.-B. Liu, Preserving qubit coherence by dynamical decoupling, \href{https://doi.org/10.1007/s11467-010-0113-8}{Front. Phys. \textbf{6}, 2 (2011)}.
\bibitem{johansson2012QuTiP2}J. R. Johansson, P. D. Nation, and F. Nori, QuTiP: An open-source Python framework for the dynamics of open quantum systems, \href{https://dx.doi.org/10.1016/j.cpc.2012.02.021}{Comp. Phys. Comm. \textbf{183}, 1760 (2012)}.
\bibitem{johansson2012QuTiP3}J. R. Johansson, P. D. Nation, and F. Nori, QuTiP 2: A Python framework for the dynamics of open quantum systems, \href{https://dx.doi.org/10.1016/j.cpc.2012.11.019}{Comp. Phys. Comm. \textbf{184}, 1234 (2013)}.
\bibitem{vlastakis2013deterministically} B. Vlastakis et al., Deterministically encoding quantum information using 100-photon Schrödinger cat states, \href{https://doi.org/10.1126/science.1243289}{Science \textbf{342}, 607 (2013)}.
\bibitem{bild2022schrodinger} M. Bild, M. Fadel, Y. Yang, U. von Lüpke, P. Martin, A. Bruno, and Y. Chu, Schrödinger cat states of a 16-microgram mechanical oscillator, \href{https://doi.org/10.1126/science.adf7553}{Science \textbf{380}, 274 (2022)}.
\ljy{\bibitem{yoshihara2017superconducting} F. Yoshihara, T. Fuse, S. Ashhab, K. Kakuyanagi, S. Saito, and K. Semba, Superconducting qubit-coscillator circuit beyond the ultrastrong-coupling regime, \href{https://doi.org/10.1038/nphys3906}{Nat. Phys. \textbf{13}, 44 (2017).}}
\ljy{\bibitem{johnson2017ultrafast} K. G. Johnson, J. D. Wong-Campos, B. Neyenhuis, J. Mizrahi, and C. Monroe, Ultrafast creation of large Schrödinger cat states of an atom, \href{https://doi.org/10.1038/s41467-017-00682-6}{Nat. Commun. \textbf{8}, 1 (2017)}.
\bibitem{hacker2019deterministic} B. Hacker, S. Welte, S. Daiss, A. Shaukat, S. Ritter, L. Li, and G. Rempe, Deterministic creation of entangled atom–light Schrödinger-cat states, \href{http://dx.doi.org/10.1038/s41566-018-0339-5}{Nat. Photonics \textbf{13} 110 (2019)}.}
\ljy{\bibitem{koch2023QRM} J. Koch, G. R. Hunanyan, T. Ockenfels, E. Rico, E. Solano, and M. Weitz, Quantum Rabi dynamics of trapped atoms far in the deep strong coupling regime, \href{https://doi.org/10.1038/s41467-023-36611-z}{Nat. Commun. \textbf{14}, 954 (2023)}.}
\cc{\bibitem{monroe1996schrodinger}  C. Monroe, D. M. Meekhof, B. E. King, and D. J. Wineland, A “Schrödinger Cat” superposition state of an atom, \href{https://doi.org/10.1126/science.272.5265.1131}{Science \textbf{272}, 1131 (1996)}.}
\ljy{\bibitem{plodzien2018rydberg} M. Płodzień, T. Sowiński, S. Kokkelmans, Simulating polaron biophysics with Rydberg atoms, \href{https://doi.org/10.1038/s41598-018-27232-4}{Sci. Rep. \textbf{8}, 9247 (2018)}.
\bibitem{stojanovic2021scalable} V. M. Stojanović, Scalable $W$-type entanglement resource in neutral-atom arrays with Rydberg-dressed resonant dipole-dipole interaction, \href{https://link.aps.org/doi/10.1103/PhysRevA.103.022410}{Phys. Rev. A \textbf{103}, 022410 (2021)}.}
\bibitem{wineland2013nobel} D. J. Wineland, Nobel lecture: superposition, entanglement, and raising Schrödinger's cat, \href{https://doi.org/10.1103/RevModPhys.85.1103}{Rev. Mod. Phys. \textbf{85}, 1103 (2013)}.
\bibitem{sychev2017enlargement} D. V. Sychev, A. E. Ulanov, A. A. Pushkina, M. W. Richards, I. A. Fedorov, and A. I. Lvovsky, Enlargement of optical Schrödinger’s cat states, \href{http://dx.doi.org/10.1038/nphoton.2017.57}{Nat. Photonics \textbf{11} 379 (2017)}.
\bibitem{liao2016generation} J.-Q. Liao, J.-F. Huang, and L. Tian, Generation of macroscopic Schrodinger-cat states in qubit-oscillator systems. \href{https://doi.org/10.1103/PhysRevA.93.033853}{Phys. Rev. A \textbf{93}, 033853 (2016)}.
\bibitem{vedral1997quantifying} \wzy{V. Vedral, M. B. Plenio, M. A. Rippin, and P. L. Knight, Quantifying Entanglement, \href{https://doi.org/10.1103/PhysRevLett.78.2275}{Phys. Rev. Lett. \textbf{78}, 2275 (1997)}.}
\bibitem{amico2008entanglement} \wzy{L. Amico, R. Fazio, A. Osterloh, and V. Vedral, Entanglement in many-body systems, \href{https://doi.org/10.1103/RevModPhys.80.517}{Rev. Mod. Phys. \textbf{80}, 517 (2008)}.}
\bibitem{bengtsson2017geometry} I. Bengtsson and K. Zyczkowski, \textit{Geometry of Quantum States: An Introduction to Quantum Entanglement} \href{https://doi.org/10.1017/CBO9780511535048}{(Cambridge University Press, Cambridge, England, 2000)}.
\cc{\bibitem{khaneja2005optimal} N. Khaneja, T. Reiss, C. Kehlet, T. Schulte-Herbrüggen, and S. J. Glaser, Optimal control of coupled spin dynamics: design of NMR pulse sequences by gradient ascent algorithms, \href{https://doi.org/10.1016/j.jmr.2004.11.004}{J. Magn. Reson. \textbf{172}, 296 (2005)}.}
\cc{\bibitem{wu2015optimal} N. Wu, A. Nanduri, and H. Rabitz, Optimal suppression of defect generation during a passage across a quantum critical point, \href{https://doi.org/10.1103/PhysRevB.91.041115}{Phys. Rev. B \textbf{91}, 041115(R) (2015)}.}
\cc{\bibitem{whitty2020eSTA}  C. Whitty, A. Kiely, and A. Ruschhaupt, Quantum control via enhanced shortcuts to adiabaticity, \href{https://link.aps.org/doi/10.1103/PhysRevResearch.2.023360}{Phys. Rev. Res. \textbf{2}, 023360 (2020)}.
\bibitem{whitty2022robust}  C. Whitty, A. Kiely, and A. Ruschhaupt, Robustness of enhanced shortcuts to adiabaticity in lattice transport, \href{https://link.aps.org/doi/10.1103/PhysRevA.105.013311}{Phys. Rev. A \textbf{105}, 013311 (2022)}.
\bibitem{sascha2022coherent}  S. H. Hauck and V. M. Stojanović, Coherent Atom Transport via Enhanced Shortcuts to Adiabaticity: Double-Well Optical Lattice, \href{https://link.aps.org/doi/10.1103/PhysRevApplied.18.014016}{Phys. Rev. Appl. \textbf{18}, 014016 (2022)}.
\bibitem{manuel2023spin}  M. Odelli, V. M. Stojanović, and A. Ruschhaupt, Spin squeezing in internal bosonic Josephson junctions via enhanced shortcuts to adiabaticity, \href{https://link.aps.org/doi/10.1103/PhysRevApplied.20.054038}{Phys. Rev. Appl. \textbf{20}, 054038 (2022)}.}


\end{thebibliography}
\end{document}